\documentstyle[twocolumn,aps,prl,epsfig]{revtex}
\def\physd#1#2#3{{Physica D}, {\bf #1}, #2 (#3)}
\def\prl#1#2#3{{Phys. Rev. Lett.}, {\bf #1}, #2 (#3)}
\def\prb#1#2#3{{Phys. Rev. B}, {\bf #1}, #2 (#3)}
\def\pre#1#2#3{{Phys. Rev. E}, {\bf #1}, #2 (#3)}
\def\prc#1#2#3{{Phys. Rep.}, {\bf #1}, #2 (#3)}

\def\beqr{\begin{eqnarray}}
\def\beq{\begin{equation}}
\def\eqn{\end{equation}\noindent}
\def\eqnr{\end{eqnarray}\noindent}

\begin{document}

\draft
\title{Collision and symmetry--breaking in the transition 
to strange nonchaotic attractors}

\author{Awadhesh Prasad and Ramakrishna Ramaswamy\thanks{rama@vsnl.com}}
\address{School of Physical Sciences, Jawaharlal Nehru University,
New Delhi 110 067, INDIA}
\author{Indubala I. Satija\thanks{isatija@sitar.gmu.edu.} and Nausheen
Shah}
\address{
Department of Physics, George Mason University, Fairfax, VA 22030, USA}
\date{\today}
\maketitle
\begin{abstract}
Strange nonchaotic attractors (SNAs) can be created due to the 
collision of an invariant curve with itself. This novel ``homoclinic''
transition to SNAs occurs in quasiperiodically driven maps which 
derive from the discrete Schr\"odinger equation for a particle in 
a quasiperiodic potential. In the classical dynamics, there is a 
transition from torus attractors to SNAs, which, in the quantum system
is
manifest as the localization transition.  This equivalence provides new
insights into a variety of properties of SNAs, including  its fractal
measure.  Further, there is a {\it symmetry breaking} associated with
the
creation of SNAs which rigorously shows that the Lyapunov exponent
is nonpositive. By considering other related driven iterative
mappings, we show that these characteristics associated with the the
appearance of SNA are robust and occur in a large class of systems.
\end{abstract}

\narrowtext

The unexpected---and fascinating---connection between strange
nonchaotic dynamics \cite{gopy} and localization phenomena
\cite{Bon,KSproc} brings together two current strands of research in
nonlinear dynamics and condensed matter physics.  The former describes
temporal dynamics converging on a fractal attractor on which the
largest Lyapunov exponent is nonpositive \cite{gopy} while the later
involves {\it exponentially} decaying wave functions. Recent work
\cite{KSloc} has shown that the fluctuations in the exponentially
decaying localized wave function are fractal, and this appears in the
classical problem as an attractor with fractal measure.  Here we
exploit this relationship further to understand the mechanism for the
transition to SNA, which is a subject of continuing interest
\cite{RMP}.

In this Letter, we show that the transition to SNAs has two unusual and
general features. Firstly, SNAs can be created by the homoclinic collision of 
invariant curves with themselves. 
Secondly, the bifurcation to SNAs, when occurring such
that the largest nontrivial Lyapunov exponent passes through zero, is
accompanied by a symmetry--breaking.  These features
provide us with a novel way to characterize and quantify the transition
to SNA. Furthermore, by considering a variety of  quasiperiodic maps, we
demonstrate that these aspects of the SNA transition are generic.

The quasiperiodically forced dynamical system under investigation here
is the Harper map \cite{KSproc},
\beq
x_{n+1} = f(x_n,\phi_n)\equiv  -[ x_n - E + 2\epsilon \cos 2\pi
\phi_n]^{-1},
\label{map}
\eqn
with the rigid--rotation dynamics $
\phi_n = n\omega +\phi_0$ giving quasiperiodic driving for irrational
$\omega$.
This map is obtained
from the Harper equation \cite{Harper},
\beq
\psi_{n+1} + \psi_{n-1} + 2\epsilon \cos[2\pi (n\omega +\phi_0)] \psi_n
=E \psi_n  \label{Harp},
\eqn
under the transformation $x_n = \psi_{n-1} / \psi_n $.  Note that the
lattice site index of the quantum problem is the time (or iteration)
index in the classical problem.  The Harper equation is a discrete
Schr\"odinger equation for a particle in a periodic potential on a
lattice.  The wavefunction at site $n$ of the the lattice is $\psi_n$,
and $E$ is the energy eigenvalue. The parameters $\epsilon$, $\omega$,
and $\phi_0$ determine the strength, periodicity and phase (relative to
the underlying lattice) of the potential.  For irrational $\omega$
(usually taken to be the golden mean, $(\sqrt{5} -1)/2$), the period of
the potential is incommensurate with the periodicity of the lattice.
For the classical map, both $\epsilon$ and $E$ are important
parameters, but the quantum problem is meaningful only when $E$ is an
eigenvalue of the system, so we limit our discussion of the classical 
system to these special values of $E$.
However, as we discuss below, this restriction can lifted when we
consider perturbations of the map which are not related to the
eigenvalue problem.  For most of our work we set $E = 0$ which is an
eigenvalue.

The Harper equation \cite{Harper} is paradigmatic in the study of localization
phenomena in quasiperiodic systems \cite{rahul}, exhibiting 
a localization transition at $\epsilon = 1$. For $\epsilon < 1$,
all eigenstates are extended and hence are characterized by an
infinite localization length, while for $\epsilon > 1$, eigenstates are
localized with localization length $\gamma^{-1} = \ln \epsilon$.
As we discuss below, the fact that the Lyapunov exponent of Harper equation
is known exactly is crucial in establishing the existence of 
SNA in the Harper map.

Of the two Lyapunov exponents for the Harper equation, that corresponding
to the $\phi$ dynamics is 0, while the other can be easily calculated
as
\beq
\label{lyap}
\lambda = \lim_{N \to \infty} {1 \over N} \sum_{i=1}^{N} y_i
\eqn
where $y_i$ is the ``stretch exponent'' defined through
\beqr
y_{i} &=& \ln |f^{\prime}(x_i)| = \ln x_{i+1}^2\nonumber \\
&=&  -2 \ln|x_i - E + 2\epsilon \cos 2\pi
\phi_i |.
\eqnr
It is easy to see that in the localized state, 
\beq
\lambda = -2 \gamma,
\eqn
and therefore, the localized wave function of the Harper equation
corresponds to an attractor with negative Lyapunov exponent for the Harper map. 

The second important point in establishing the existence of SNA in Harper
equation stems from the fact that
the fluctuations about the localized wave function in the Harper equation are
fractal. This result, based on renormalization studies\cite{KSloc} of the Harper equation,
suggests that the corresponding attractor
in the Harper map has a
fractal measure and hence is an SNA. Furthermore, a perturbative
argument starting from the strong coupling limit provides a 
rigorous proof for the existence of SNA for $E=0$ \cite{KSproc}, making the
Harper mapping one of the few systems where the 
existence of SNA is well established.

\begin{figure}
\epsfig{bbllx=10pt,bblly=40pt,bburx=515pt,bbury=460pt,figure=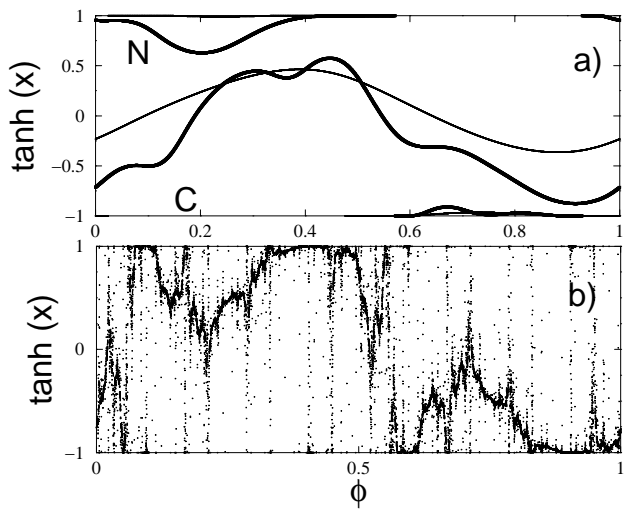,width=8.5cm}
\caption{(a) Invariant curves for the Harper map with $E = 0$ 
at $\epsilon=0.3$ (thin line) and $\epsilon=0.65$ (thick line) with
initial conditions $x_0=0.5$ and $\phi_0=0.4$. (We plot
$\tanh(x)$ rather than $x$ on the ordinate so that the entire
range of $x$ can be depicted.)  For different initial
conditions we will get different curves.  As $\epsilon$ is increased,
the two branches of a given curve approach each other and eventually
collide as $\epsilon \to 1$ as indeed do all the other curves as well.
The Central and Noncentral branches of these curves are marked C and N
respectively.
(b) The SNA that is born at $\epsilon = 1$. }
\label{fig2}
\end{figure}

We now discuss the scenario for the formation of SNAs in this 
system when $E=0$. 
For $\epsilon < 1$, the phase space is foliated by invariant
curves, each parametrized by the initial conditions. 
It is important to note that for $\epsilon < 1$ there are no attractors
in the system since all the curves are neutrally stable. However,
at $\epsilon =1$, trajectories converge on
an attractor. The convergence is power-law and hence the
Lyapunov exponent is zero:  we can characterize this via a
power-law exponent
\beq
\beta= \lim_{N\to \infty} {1\over \ln N } \ln \prod_{i=0}^{N-1}
\exp y_i
\eqn  the transition from a family of invariant tori to
an attractor being signaled by a non-zero value of $\beta$.

The transition from an invariant curve to the attractor can
be described as a collision phenomenon as we discuss below.
For $\epsilon < 1$, the invariant
curves have two branches [see Fig.~1(a)] deriving from the fact that
for $\epsilon = 0$, the map does not have a period--1 fixed point for
real $x$ but has instead a period--2 orbit. 
As $\epsilon \to 1$, the two branches approach each other
and {\it collide} at $\epsilon = 1$, the point of collision being
a singularity. Since the dynamics in $\phi$ is ergodic, the collision
occurs at a dense set of points. Furthermore, this happens for
each invariant curve, and in effect all invariant curves approach 
each other and collide at $\epsilon = 1$, forming an attractor [see Fig.~1(b)].
We quantify this collision by demonstrating that as
$\epsilon \to 1$, the distance $d$ between the two branches goes to
zero as a power--law.

When the quasiperiodic forcing frequency $\omega$ is the golden
mean ratio, the distance between the two branches of an invariant 
curve can be calculated by first noting that
a point $(x_i,\phi_i)$ and its successive Fibonacci iterates,
$(x_{i+F_k}, \phi_{i+F_k})$, where $F_k$ is the $k'$th Fibonacci number,
are closely spaced in $\phi$ \cite{general}. If the two branches of the 
invariant curve are labeled C (for central) and N (for noncentral) 
[see Fig.~1(a)], the 
sequence of Fibonacci iterates follows the symbolic coding CCNCCNCCNCCN 
$\ldots$ or NNCNNCNNCNNC$\ldots$. This follows from the fact that the
Fibonacci numbers are successively even,
odd, odd, even, odd, odd,$\ldots$. Thus, if $k$ is chosen appropriately, 
such that $F_k$ is even and $F_{k+1}$ is odd (or vice-versa),
\beq
d_k(i) = \vert x_{i+F_k} - x_{i+F_{k+1}}\vert
\eqn
measures the approximate vertical distance between the curves at 
$(x_i,\phi_i)$. 
Minimizing this distance along the invariant curve, we find that the
closest approach of the two branches decreases as a power, 
\beq
d= \min[ \lim_{k \to \infty} d_k(i)] \sim (1-\epsilon)^{\delta}.
\eqn
Our results, given in Fig.~2, provide a {\it quantitative} 
characterization of the transition to SNA in this system.

For eigenvalues other than $E=0$, the scenario for SNA 
formation may be different.
When the eigenvalue $E$ is at the band--edge, the 
SNAs appear to be formed via the fractalization
route, namely by gradually wrinkling and forming a fractal
\cite{fractal}. The reason for this difference can be traced 
to the simple fact
that unlike the $E=0$ case, below $\epsilon = 1$
the invariant curve for the minimum eigenvalue has a single branch which
originates from a fixed point for $\epsilon = 0$.

The self--collision of invariant curves to form SNAs is a general
mechanism. Consider a family of maps,
\begin{eqnarray}
\nonumber
x_{i+1} &=& -[ x_i +\alpha x_i^{\nu} + 2\epsilon \cos 2\pi \phi_i]^{-1}
\\
\phi_{i+1} &=&\phi_i + \omega {~~~~~\mbox{mod}}~1\label{map2v}
\label{perturb}
\end{eqnarray}\noindent
which bear no relation to an eigenvalue problem.  For $\nu$ an odd
integer, the above map is invertible and hence does not have any
chaotic attractors.  Numerical results for $\nu=3$ show that in these
perturbed maps, an SNA is also born after the attractor collides with
itself.  Similar results are obtained for other polynomial or
sinusoidal perturbations.

\begin{figure}
\epsfig{bbllx=140pt,bblly=420pt,bburx=515pt,bbury=660pt,figure=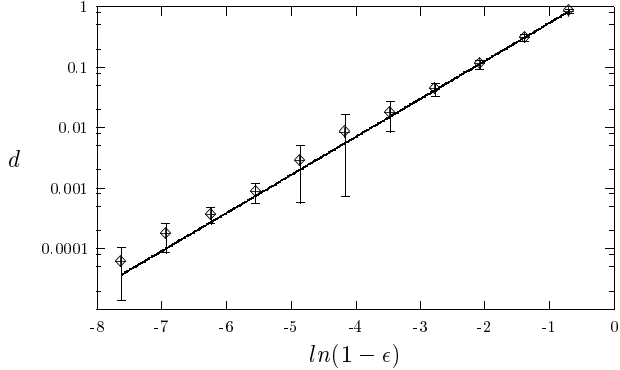,width=8.5cm}
\caption{The minimum vertical distance $d$ between the branches of invariant
curves in the Harper map as a function of the parameter $(1 - \epsilon)$,
for the eigenvalue $E = 0$ and for the Fibonacci iterates (see the text)
$F_k$ and $F_{k+1}$ with $k = 23$. Since this quantity depends
on the particular curve chosen, an average is taken over 12
different invariant curves, which gives the average behaviour and the 
variance. The resulting exponent [the solid line is a 
fit to a power--law; see Eq.~(8)] is $\delta \approx 1.4$.}
\label{fig3}
\end{figure}

A more fundamental characteristic of this route to SNAs is a dynamical
symmetry--breaking.
Although the dynamics is nontrivial for the variable $x$, the Lyapunov
exponent is {\it exactly} zero for $\epsilon < 1$.  To understand this
from a dynamical point of view, we first note that for finite times
along a trajectory, the local expansion and contraction rates vary. It
turns out that a meaningful way to understand the role of the parameter
$\epsilon$ is to study the return--map for the stretch exponents, 
\beq
\label{lyapmap}
y_{i+1} = -2\ln |\mbox{sgn}(x_i) \exp (y_i/2) - E + 2 \epsilon \cos
2\pi\phi_i|.
\eqn
Shown in Fig.~3(a) is the above map for $E = 0$ and $\epsilon =
0.5$. There is a reflection symmetry evident, namely $(x,y \to -y,-x)$
although this symmetry is not easy to see directly in the mapping,
Eq.~(\ref{lyapmap}) itself owing to the quasiperiodic nature of the
dynamical equations. However, as a consequence of the symmetry, the
positive and the negative terms cancel exactly in Eq.~(\ref{lyap}),
giving a zero Lyapunov exponent.  All finite sums of the stretch
exponents, namely the finite--time Lyapunov exponents \cite{ftl} also
share the same symmetry features.

\begin{figure}
\epsfig{bbllx=10pt,bblly=0pt,bburx=330pt,bbury=730pt,figure=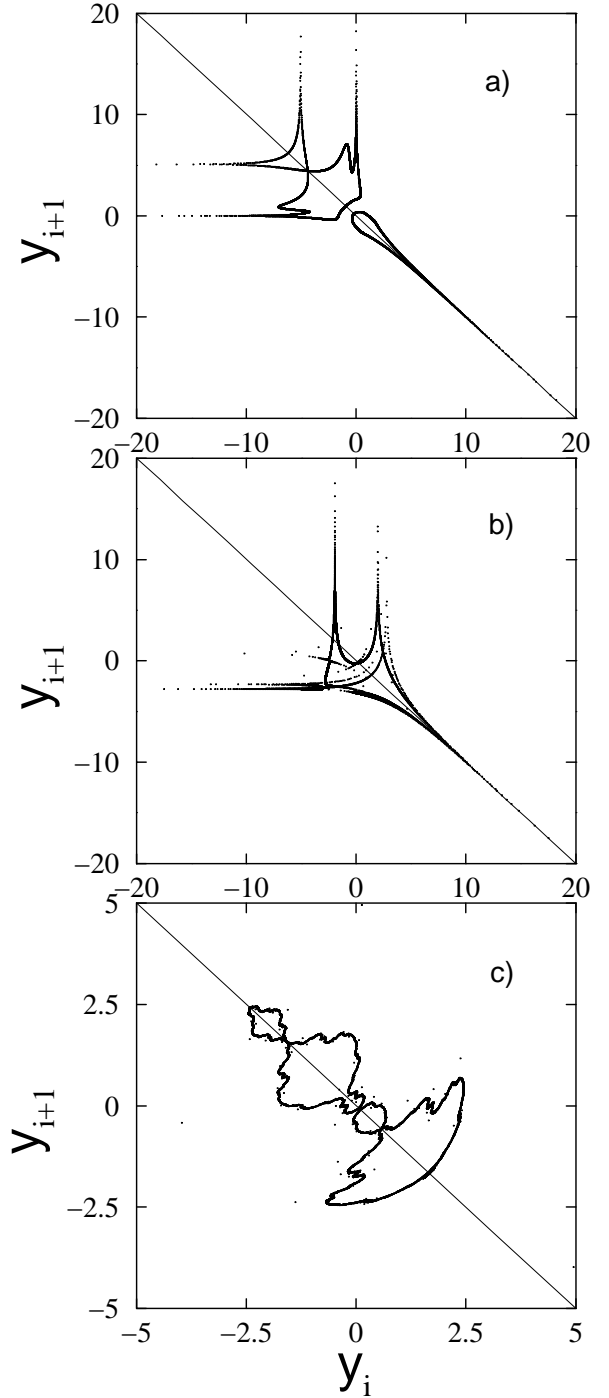,width=8.5cm}
\caption{ The return map for the stretch exponents, 
Eq.~(\ref{lyapmap}) (a) for $E=0, \epsilon=
0.5$, with initial conditions, $x_0=0.2, \phi_0 = 0.7,$ 
(b) for  $E=0, \epsilon=2$, and  (c) at the band edge, 
$E = E_m =-2.597515185\ldots, \epsilon = 1$. }
\label{fig1}
\end{figure}

This symmetry is maintained for $0 < \epsilon \le 1$, above which
this symmetry is broken [Fig.~3(b)]. When the negative
stretch exponents exceed the positive ones, the Lyapunov exponent
$\lambda$ therefore becomes negative; coupled with the fact that the
attractor has a dense set of singularities \cite{KSproc}, this
rigorously confirms the existence of strange nonchaotic dynamics.

Symmetry--breaking appears to be operative in a large class of systems,
including the mapping where SNAs were first shown to exist \cite{gopy},
viz. $x_{n+1}= 2\epsilon \cos 2\pi \phi_n \tanh x_n$, and similar
systems where the transition to SNA is via the blowout bifurcation
\cite{blowout}. In all these instances, the largest Lyapunov exponent
goes through zero when the SNA is born.

When the eigenvalue $E$ differs from 0, say at the band--edge,
the attractor in the localized state is also a SNA which is
born at $\epsilon = 1$, with zero Lyapunov exponent. Again [see Fig.~3(c)]
there is the symmetry in the return map for the stretch
exponents which is broken for $\epsilon > 1$.

In summary, our work shows that the fractal measure of the trajectory
has its origin in the homoclinic collisions of an invariant curve with
itself.  This characterization of the transition to SNAs can be
quantified, and  may serve as a useful scenario for the 
appearance of SNAs in a variety of nonlinear dissipative systems.  

Furthermore, we demonstrate that the transition from an invariant curve to 
a SNA proceeds via a symmetry--breaking. A zero value for the Lyapunov 
exponent of a system
can arise in a number of ways, and the present instance, namely the
exact cancelation of expanding and contracting terms is very special.
(There is similar symmetry breaking at all period--doubling
bifurcations in such systems as well, but these points are of
measure zero.) It is conceivable that there are
more complex symmetries in other systems which similarly lead to a zero
value for the Lyapunov exponent.  The significance of this symmetry and
its breaking in the corresponding quantum problem may be an important
question in characterizing the localization transition itself.  

There are numerous lattice models
exhibiting localization in aperiodic potentials\cite{Soko},
including the quantum kicked rotor\cite{Fishman,SB}. The 
corresponding derived  aperiodic mappings are worthy of
further study and might well extend the subject of SNA to
systems beyond quasiperiodically driven maps.
In addition, there are interesting open questions
regarding localization and its absence in quasiperiodic potentials with
discrete steps \cite{KSdimer}. It is conceivable that this type of
mapping of the quantum problem onto the classical problem may provide
better understanding of localization phenomena.

{\sc ACKNOWLEDGMENT:} We would like to thank U. Feudel, J. Ketoja, and J.
Stark for various illuminating discussions during the seminar ``Beyond
Quasiperiodicity'' where this work was started.  We would also like to
acknowledge the hospitality of the Max Planck Institute for Complex
Systems, Dresden.  RR is supported by the Department of Science and
Technology, India, and AP by the CSIR. The research of IIS is supported
by grant DMR~097535 from the National Science Foundation.


\begin{references}
\bibitem{general} The Fibonacci numbers are recursively defined
through $F_{k+1}=F_k+F_{k-1}$ with $F_{-1}=0, F_0=1$, and $\lim_{k \to \infty} F_k/F_{k+1} 
\to (\sqrt 5 -1 )/2$, the golden mean ratio. If the driving 
frequency is some other irrational number, then
it will be necessary to consider different sequences of iterations,
which will depend on the successive rational approximants to the 
frequency.
\end{references}
\end{document}